\documentclass[twocolumn,secnumarabic,showpacs,showkeys,amssymb, amsmath,nobibnotes, aps, pre]{revtex4-1}

\usepackage{graphicx}

\begin{document}


\title{Equilibrium distributions and relaxation times in gas-like 
economic models: an analytical derivation}

\author{Xavier Calbet}
\email{xcalbet@googlemail.com}
\affiliation{
BIFI,\\
Universidad de Zaragoza,\\
E-50009 Zaragoza, Spain.\\
}

\author{Jos\'e-Luis L\'opez}
\email{jl.lopez@unavarra.es}
\affiliation{
Dept. of Mathematical and Informatics Engineering,\\
Universidad P\'ublica de Navarra,\\
E-31006 Pamplona, Spain.\\
}

\author{Ricardo L\'opez-Ruiz}
\email{rilopez@unizar.es}
\affiliation{
DIIS and BIFI,\\
Universidad de Zaragoza,\\
E-50009 Zaragoza, Spain.\\
}

\date{\today}

\begin{abstract}
A step by step procedure to derive analytically 
the exact dynamical evolution equations of the probability density
functions (PDF) of well known kinetic wealth exchange
economic models is shown. This technique gives a dynamical insight
into the evolution of the PDF, e.g., 
allowing the calculation of its relaxation times.
Their equilibrium PDFs can also be calculated
by finding its stationary solutions.
This gives as a result an integro-differential
equation, which can be solved analytically in some cases
and numerically in others. This should provide some guidance
into the type of probability density functions 
that can be derived from particular economic
agent exchange rules, or for that matter, any other kinetic model of gases
with particular collision physics.
\end{abstract}

\pacs{89.75.Fb, 05.45.-a, 02.50.-r, 05.70.-a}
\keywords{econophysics, economic models, molecular dynamics}
\maketitle

\section{Introduction}

The aim of mechanical statistics is to describe systems macroscopically
based on the microscopic description of interactions of particles within
the system. By defining the microscopic individual interaction
of a collection of particles we can describe the system
macroscopically by means of the Probability Density Function (PDF).
A particular microscopic interaction will give rise to a definite
PDF. And vice versa, a given PDF can come only from a small
set of specific particle interactions.
In this respect, the macroscopic PDF is also providing us with
information about the microscopic interactions.
A classical example is the Maxwell-Boltzmann distribution,
which can be obtained as a solution of the Boltzmann integro-differential
equation, which he proposed to explain the evolution of the PDF
for a dilute gas.

Relatively recently, there has been growing interest in
reproducing the PDF of money in a real economic system
by molecular dynamics simulations \cite{chakraborti00},
\cite{dragulescu00}, \cite{chakraborti02}, \cite{hayes02},
\cite{chatterjee03}, \cite{das03}, \cite{lopez-ruiz09}. In these statistical
models, agents are allowed to exchange money following an exchange
rule. PDFs from real economies follow Gibb's exponential functions
or Pareto's laws \cite{yakovenko09}.
Reproducing Gibb's exponential functions with molecular dynamics
simulations has proven simple by using a simple money exchange rule
that conserves the total quantity of money \cite{dragulescu00}.
Pareto's law distributions still remain a challenge, although
some results do approximate it \cite{levy96},\cite{estevez08},
\cite{pellicer-lostao09}, \cite{chakraborti09}.

In this paper, we will show how to derive analytically
the dynamical evolution equation of the PDF 
from the interaction rules of the agents.
This process is very similar to the derivation of the
Boltzmann integro-differential equation from the basic
particle collision physics. 
Let us recall that this latter system has a 
Maxwell-Boltzmann PDF in equilibrium.
Repetowicz et al. \cite{repetowicz04} have derived, by using
mean-field approximation, the first moments
of some of these economic models based on the formal solutions
to the nonlinear Boltzmann equations derived by
Ernst \cite{ernst81}.
Lallouache et al. \cite{lallouache10}
also calculate the moments of the
random gas-like economic model with savings 
and its steady state solution.
D\"uring et al. derive the moments of some economic models
and their relaxation times \cite{during08}.
We will demonstrate a simple step by step technique,
different from the ones above,
to derive the integro-differential equations of three economic models
present in the reviews by Patriarca et al. \cite{patriarca09},
Chatterjee and Chakrabarti \cite{chatterjee07} or 
Yakovenko and Rosser \cite{yakovenko09b}. 
Some of these systems exhibit exponential
PDFs as their steady state solutions. 
Other systems, which seem to follow Gamma distributions \cite{patriarca04}
at their steady state, are in fact only approximated by them and do not 
follow them exactly \cite{lallouache10}. We will show that
some systems, in some particular cases, deviate appreciably from
these Gamma distributions. We will derive the analytical formulas
to calculate these distributions and we will
compare these PDFs with Gamma distributions and molecular
dynamics simulations.
This technique should shed some light into the relationships
between PDFs and the microscopic interactions of particles,
and in particular, into the derivation of a Pareto type
distribution from molecular dynamics simulations.

We also present, to our knowledge, for the first time,
the dynamical equations of the evolution of the PDFs
for these systems.
This will allow us to study these systems not only
from the point of view of their
stationary PDFs, but also how they evolve in time.
This area of research has often been neglected in the past,
but it is key if we want to model the economy adequately
by describing its evolution in time and not only
in a static way.
As an example, we will show how to derive the
relaxation times of these systems.

\section{Pure random gas-like economic model}

\begin{figure}
\includegraphics[angle=0,width=0.9\columnwidth]{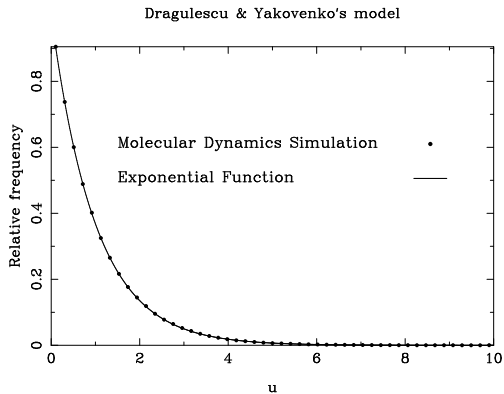}
\caption{Dragulescu and Yakovenko's economic model \cite{dragulescu00} 
which gives an exponential as the steady state PDF.
Dots are the result of the molecular dynamics numerical simulations,
and solid line is the exponential function which approximates the steady
state PDF (Eq. \ref{eq:exponential}).
For these calculations, we take $<u>=1$.}
\label{fig:exponential}
\end{figure}

We will introduce the technique directly with an example.
In this case, we take the mapping introduced by Dragulescu 
and Yakovenko \cite{dragulescu00} to model the flow and distribution 
of money. This example is one of the simplest money exchange rules
in which money is conserved in any transaction.

Assuming that we have $N$ agents trading with each other,
the mapping that describes this statistical model is,

\begin{eqnarray}
\label{eq:gibbs}
u'_i &=& \epsilon \; (u_i + u_j), \nonumber \\
u'_j &=& (1 - \epsilon) (u_i + u_j),\nonumber\\
i , j &=& 1 \ldots N,
\end{eqnarray}

where $\epsilon$ is a random generated number in the interval $(0,1)$, 
$u_i$ and $u_j$ are the initial money (or energy) and $u'_i$ and $u'_j$ are 
the final ones of agents $i$ and $j$, respectively, where the pair of 
agents $(i,j)$ is also randomly chosen for each transaction. 
These quantities, $u_i$, will always be positive. 
The steady state distribution, $f_{eq}$, obtained with numerical simulations 
of this system is shown in Fig. \ref{fig:exponential} (dotted line).
It can be easily verified that it is an exponential 
or Gibbs distribution,

\begin{equation}
\label{eq:exponential}
f_{eq}(u)=\beta \exp(-\beta u),
\end{equation}

where $f(u) {\mathrm d}u$ denotes the PDF, or probability
of finding an agent with money (or energy) between $u$ and $u + {\mathrm d}u$. 
This PDF is normalized and the mean energy per particle then turns out to be

\begin{equation}
 \langle u \rangle = \frac{1}{\beta}.
\end{equation}

Let us now derive this function analytically.
To do this, it is not only important
the money exchange mapping (Eqs. \ref{eq:gibbs}) in the
determination of the PDF, but also 
the selection rule to choose which particles do interact at each 
time step. Therefore, we will explicitly
describe the detailed and complete algorithm,

\begin{enumerate}

\item
\label{gibbs:rl:1}
A pair of different agents are randomly selected in the 
system from a uniform distribution
in the interval $[1,N]$. This is the pair $(i,j)$ in Eqs. \ref{eq:gibbs}.

\item
\label{gibbs:rl:3}
A random number $\epsilon$ between 0 and 1 is
generated from a uniform distribution.

\item
\label{gibbs:rl:4}
Exchange rules of Eqs. \ref{eq:gibbs} are finally applied.

\item
The application of all of the above
``rules'' will be denoted as a ``step''. 
These steps are successively and indefinitely repeated.

\end{enumerate}

We now derive the analytical expression for the PDF variation
after each step, $\partial f(u) / \partial t$, for a given money $u$.
Note that with this notation we assume that the time unit is
one step.
As explained in the next paragraphs in more detail, 
it can be straightforwardly seen that this variation comes 
on one side from the probability that agent $u$ has to be selected 
for a particular exchange, i.e., that $u_i$ or $u_j$ take the value $u$,
and from another side from 
the probability that the result after the trading is $u$,
i.e., that $u'_i$ or $u'_j$ give the value $u$. We can then write,

\begin{eqnarray}
\frac{\partial f(u)}{\partial t} & = & \left[ \frac{\partial f(u_i)}{\partial t} + \frac{\partial f(u'_i)}{\partial t} \right. + \nonumber \\ 
 & & \left. \frac{\partial f(u_j)}{\partial t} + \frac{\partial f(u'_j)}{\partial t} \right]_{u_i=u_j=u'_i=u'_j=u},
\label{eq:formula1}
\end{eqnarray}

where all terms are maintained separated in the case that the
trading rule is not symmetric in the indices $(i,j)$
before or after the interaction.

Let us see the detailed explanation for the present example.
In rule \ref{gibbs:rl:1}, one particular agent from the whole 
population $N$ is selected. The probability  that this particular
agent is in the $[u_i , u_i + {\mathrm d}u_i]$
range is proportional to the number of agents in that range,
then the PDF for $u_i$ will be proportionally depleted  
by the quantity $f(u_i) {\mathrm d}u_i$, that is,

\begin{equation}
\label{eq:gibbs:deltafu1}
\frac{\partial f(u_i)}{\partial t} \sim - f(u_i).
\end{equation}

Rule \ref{gibbs:rl:1} will also deplete the PDF 
for $u_j$ in a similar way, that is, $\partial f(u_j)/\partial t \sim - f(u_j)$.

Rules \ref{gibbs:rl:3} and \ref{gibbs:rl:4} imply that,
since $\epsilon$ is between 0 and 1, and according
to Eqs. \ref{eq:gibbs}, the net result verifies

\begin{equation}
\label{eq:gibbs:ineq}
 0 < u'_i < u_i + u_j,
\end{equation}
 
where $u'_i$ is equally distributed in the $[0, u_i + u_j]$ interval. 
This implies that the PDF for $u'_i$ is increased proportionally to the number
of agents with $u_i$ and $u_j$ and inversely to their total money $u_i+u_j$,
that is, to the rate 
$f(u_i)  {\mathrm d}u_i f(u_j) {\mathrm d}u_j /(u_i + u_j)$.
Then, the total variation $\partial f(u'_i)/\partial t$ will be obtained
by integrating amongst all the possible
values of $u_i$ and $u_j$ that give rise to the result $u'_i$.
In order to find the limits of integration in a simple way, 
we make a change of variables replacing $(u_i,u_j)$ by $(u_i, U)$, 
where 

\begin{equation}
\label{eq:defu}
U \equiv u_i + u_j.
\end{equation}
 
Equation \ref{eq:gibbs:ineq} is then transformed into $0 < u'_i < U$,
which forces the integration limits of $U$ to be between $u'_i$ and $\infty$.
If $U$ is now fixed, and using its definition from Eq. \ref{eq:defu}, 
the integration limits of $u_i$ have to be between $0$ and $U$.
With these new variables, the expression for $\partial f(u'_i)/\partial t$ is 
finally obtained

\begin{equation}
\label{eq:gibbs:deltafup1_2}
 \frac{\partial f(u'_i)}{\partial t} \sim \int_{u'_i}^\infty \mathrm{d} U \int_{0}^U \mathrm{d} u_i 
\frac{f(u_i) f(U - u_i)}{U}.
\end{equation}

By symmetry of Eqs. \ref{eq:gibbs}, a similar result
for $\partial f(u'_j)/\partial t$ is just obtained by substituting 
$u'_i$ for $u'_j$ in Eq. \ref{eq:gibbs:deltafup1_2}.

As indicated by Eq. \ref{eq:formula1}, to obtain the final result for 
a generic value $u$ we combine the decreasing term of Eq. \ref{eq:gibbs:deltafu1}
with the positive contribution of Eq. \ref{eq:gibbs:deltafup1_2} to get

\begin{equation}
 N \frac{\partial f(u)}{\partial t} = -  2 f(u) +  2 \int_{u}^\infty \mathrm{d} U \int_{0}^U \mathrm{d} u_1 
\frac{f(u_1) f(U - u_1)}{U},
\end{equation}

where $N$ is the normalization factor for $\partial f(u)/\partial t$.

To derive the steady state solution we just set $\partial f(u)/\partial t = 0$ to finally obtain

\begin{equation}
\label{eq:gibbs:final}
 f_{eq}(u) = \int_{u}^\infty \mathrm{d} U \int_{0}^U \mathrm{d} u_1 
\frac{f_{eq}(u_1) f_{eq}(U - u_1)}{U}.
\end{equation}

It is now easy to verify that the normalized exponential distribution 
from Eq. \ref{eq:exponential} satisfies this last equation.
This analytical solution is plotted in Fig. \ref{fig:exponential},
where it is verified that it is very much 
in agreement with the numerical simulations.

We shall now derive the relaxation time of this system.
We will assume that we start with a PDF that is different from the
equilibrium one (Eq. \ref{eq:exponential}), but close to it.
If we substitute this PDF into
the right hand side of Eq. \ref{eq:gibbs:final},
we can verify (not shown here) that we obtain a very good approximation
to the exponential equilibrium PDF (Eq. \ref{eq:exponential}).
The evolution equation close to the equilibrium is then simplified to,

\begin{equation}
 N \frac{\partial f(u)}{\partial t} \simeq -  2 f(u) +  2 f_{eq}(u).
\end{equation}

From here, we can immediately see that the relaxation time of the
system is

\begin{equation}
\tau = \frac{N}{2}
\end{equation}

collisions.

\section{Random gas-like economic model with saving}

\begin{figure}
\includegraphics[angle=0,width=0.9\columnwidth]{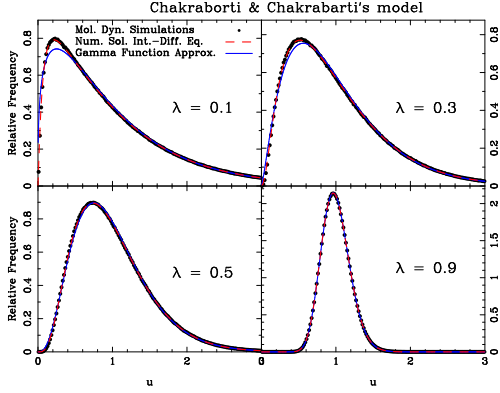}
\caption{Chakraborti and Chakrabarti's economic model \cite{chakraborti00},
which gives approximately a Gamma function as the steady state PDF, plotted 
for different values of the saving parameter, $\lambda$.
Dots are the result of the molecular dynamics numerical simulations,
blue solid line is the Gamma function which approximates the steady
state PDF (Eq. \ref{eq:gamma2}) and
red dashed line is the PDF which is the solution
to the integro-differential equation \ref{eq:gamma2:final}.
For these calculations, we take $<u>=1$.}
\label{fig:chakra}
\end{figure}

We will now explore the gas-like economic model with saving 
introduced by Chakraborti and Chakrabarti \cite{chakraborti00}.
The trading rules in this model are given by a random mapping that
conserves the amount of money in each exchange, and in which
each agent saves a fixed fraction, $\lambda$, of the money he owns
before the transaction. More precisely, money is exchanged
in the form,

\begin{eqnarray}
\label{eq:chakra}
u'_i &=&  \lambda \; u_i + \epsilon \; ( 1 - \lambda ) \; (u_i + u_j), \nonumber \\
u'_j &=&  \lambda \; u_j + (1 - \epsilon) \; ( 1 - \lambda ) \; (u_i + u_j),\nonumber\\
i , j &=& 1 \ldots N.
\end{eqnarray}

The steady state distribution obtained by numerical
simulations of this system is shown in Fig. \ref{fig:chakra} (dotted lines).
It is verified \cite{patriarca09} that it can be approximated
by a Gamma distribution,

\begin{equation}
\label{eq:gamma2}
f_{eq}(u) \simeq a u^{n - 1} \exp \left(-n u/\langle u \rangle \right),
\end{equation}

where $\langle u \rangle$ is the mean wealth of the multi-agent system, 
and $n$ is given by

\begin{equation}
 n = \frac{1 + 2 \lambda}{1 - \lambda}.
\end{equation}

The factor $a$ is found after the normalization of the PDF,

\begin{equation}
 a = \frac{1}{\Gamma(n)} \left( \frac{n}{\langle u \rangle} \right)^n.
 \label{eq:Gamma_a}
\end{equation}

This Gamma function is plotted as a blue solid line in Fig. \ref{fig:chakra}.
It can be observed that it adjusts relatively well to the numerical simulations.

To derive the analytical expression of the steady state PDF we have to look at
the precise rules for this mapping:

\begin{enumerate}

\item
\label{gamma2:rl:1}
A pair of different agents of the system are randomly selected from a uniform distribution
in the interval $[1,N]$. This will be the trading pair $(i,j)$ in Eqs. \ref{eq:chakra}.

\item
\label{gamma2:rl:3}
A random number $\epsilon$ between 0 and 1 is
generated from a uniform distribution.

\item
The saving parameter $\lambda$ is maintained constant
through the whole process.

\item
\label{gamma2:rl:4}
Exchange rules of Eqs. \ref{eq:chakra} are finally applied.

\item
The application of all of the above
``rules'' will be denoted as a ``step''. 
These steps are successively and indefinitely repeated.

\end{enumerate}

Let us derive now in an analytical way the PDF for this system.
As in the previous model, rule \ref{gamma2:rl:1} will deplete 
the density function $f(u)$ in $u_i$ by a quantity $\partial f(u_i)/\partial t$
given by

\begin{equation}
\label{eq:gamma2:deltafu1}
\frac{\partial f(u_i)}{\partial t} \sim - f(u_i),
\end{equation}
and similarly, $\partial f(u_j)/\partial t \sim - f(u_j)$.

The mapping equations (Eqs. \ref{eq:chakra}) lead 
to the following inequalities

\begin{equation}
 \lambda u_i < u'_i < \lambda u_i + (1-\lambda)U,
 \label{eq:limits}
\end{equation}

where $u'_i$, by using again $U = u_i + u_j$, is equally distributed 
in the $[\lambda u_i, \lambda u_i + (1 - \lambda) U]$ interval. 
This implies that the PDF for $u'_i$ is increased proportionally to the number
of agents with $u_i$ and $u_j$ and inversely to the length of the interval,
$(1 - \lambda) U$, where it is distributed, that is, to the rate 
$f(u_i) {\mathrm d}u_i f(u_j) {\mathrm d}u_j /((1 - \lambda) U)$.
Then, the total variation $\partial f(u'_i)/\partial t$ will be obtained
by integrating amongst all the possible
values of $u_i$ and $u_j$ able to give rise to the result $u'_i$.
The limits of integration can be easily found from Eq. \ref{eq:limits}:

\begin{eqnarray}
 u_i & < & u'_i/\lambda,\nonumber \\
 u_i & > & (u'_i - (1 - \lambda)U)/\lambda
\end{eqnarray}

As in the former section for the Dragulescu and Yakovenko model \cite{dragulescu00},
we obtain the result:

\begin{eqnarray}
\label{eq:gamma2:deltafup1}
\frac{\partial f(u'_i)}{\partial t} \sim & & \nonumber \\
\int_{u'_i}^\infty \mathrm{d} U
\int_{\max\left[\frac{u'_i-(1-\lambda)U}{\lambda},\,0\right]}^{\min[u'_i/\lambda,\,U]} 
\mathrm{d} u_i \frac{f(u_i) f(U - u_i)}{( 1 - \lambda) U}. & &
\end{eqnarray}

A similar formula can be derived for the $u'_j$ variable. 
Combining these results (Eqs. \ref{eq:gamma2:deltafu1} and \ref{eq:gamma2:deltafup1}), 
the final expression for the PDF variation $\partial f(u)/\partial t$ in a generic value $u$
is derived:

\begin{eqnarray}
\label{eq:gamma2:evolution}
N \frac{\partial f(u)}{\partial t} = \nonumber \\
- 2 f(u) + \nonumber \\
 2 \int_{u}^\infty \mathrm{d} U \int_{\max\left[\frac{u-(1-\lambda)U}
{\lambda},0\right]}^{\min\left[u/\lambda,U\right]} \mathrm{d} u_1 
\frac{f(u_1) f(U - u_1)}{( 1 - \lambda) U},
\end{eqnarray}

where $N$ is again the normalization factor of $\partial f(u)/\partial t$.

To find the steady state solution, we just set $\partial f(u)/\partial t = 0$ to finally get

\begin{equation}
\label{eq:gamma2:final}
 f_{eq}(u) = \int_{u}^\infty \mathrm{d} U \int_{\max\left[\frac{u-(1-\lambda)U}
 {\lambda},0\right]}^{\min\left[u/\lambda,U\right]} \mathrm{d} u_1 
\frac{f_{eq}(u_1) f_{eq}(U - u_1)}{( 1 - \lambda) U}.
\end{equation}

Eq. \ref{eq:gamma2:final} can be solved iteratively 
by feeding as first approximation the gamma function 
from Eq. \ref{eq:gamma2} in its right hand side
and solving the integrals
numerically to give a second order PDF approximation in the left
hand side. 
We need only to iterate once to get a better approximation 
than the first one.
Results for different parameters of $\lambda$ are shown 
in Fig. \ref{fig:chakra} as a dashed red line, together with molecular
dynamics simulations and the first approximation itself (Eq. \ref{eq:gamma2}).
It can be verified how the the Gamma distribution deviates
substantially from the molecular dynamics simulations 
for the cases with a low value of $\lambda$. On the contrary, 
the numerical solution of the integro-differential equation 
(Eq. \ref{eq:gamma2:final}) matches very well 
the molecular dynamics simulations.
A similar expression of this steady state solution, derived
in a different way, was found
by Lallouache et al. (Eq. 34 in \cite{lallouache10}).
 
As before, we can now calculate the relaxation time
by starting with a PDF close to equilibrium.
In this case we will start with a Gamma function (Eq. \ref{eq:gamma2}),
which we know is not the distribution at the equilibrium, but it is close
to it. As we have seen(Fig.\ref{fig:chakra}), 
if we place this PDF on the right hand side
of Eq. \ref{eq:gamma2:final} we obtain a very good approximation
of the equilibrium distribution. That leaves the evolution equation
(Eq. \ref{eq:gamma2:evolution})
close to equilibrium simplified to,

\begin{equation}
 N \frac{\partial f(u)}{\partial t} \simeq -  2 f(u) +  2 f_{eq}(u),
\end{equation}

which again leads us to a relaxaion time of $\tau = N/2$
collisions.

\section{Asymmetric random gas-like economic model}

\begin{figure}
\includegraphics[angle=0,width=0.9\columnwidth]{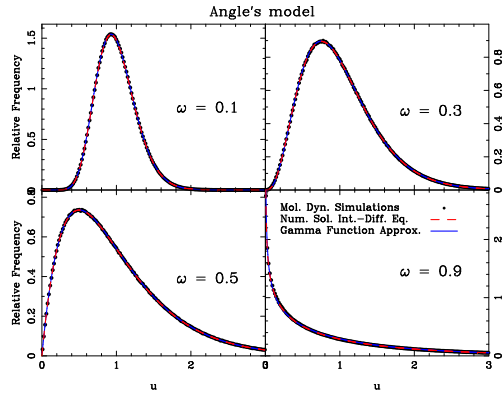}
\caption{Angle-like's economic model \cite{angle83}
which gives approximately a Gamma function
as the steady state PDF plotted for different values of $\omega$.
Dots are the result of the molecular dynamics numerical simulations,
blue solid line is the Gamma function which approximates the steady
state PDF (Eq. \ref{eq:gamma2}), 
red dashed line is the PDF which is the solution
to the integro-differential equation \ref{eq:gamma1:final}.
For these calculations, we take $<u>=1$.}
\label{fig:gamma1}
\end{figure}

We will now explore a mapping given in \cite{patriarca09}
as a modification of a model introduced by Angle \cite{angle83}.
This mapping has a parameter $\omega$ which graduates the amount of exchange
of money between interacting agents.
We will only explore the 
simplest version of all these possible Angle-like models,
where one agent gives money to another
one regulated by $\omega$ and a random number.
It is an asymmetric model because in this model one of the agents is chosen as a winner
and the other one as a loser.
Other more sophisticated versions could as well be explored with this method.
The precise rules to follow for this mapping are,

\begin{enumerate}

\item
\label{gamma1:rl:1}
We select randomly using a uniform distribution two different 
agents $(i,j)$.

\item
\label{gamma1:rl:3}
We obtain a random number between 0 and 1, $\epsilon$,
generated from a uniform distribution.

\item
The exchange parameter $\omega$ is maintained constant
through the whole process.

\item
\label{gamma1:rl:4}
Agent $u_j$ will now give some money to agent $u_i$ according to
the following mapping,

\begin{eqnarray}
\label{eq:gamma1}
 u'_i &=& u_i + \epsilon \; \omega \; u_j, \nonumber \\
 u'_j &=& u_j - \epsilon \; \omega \; u_j,\nonumber\\
i , j &=& 1 \ldots N.
\end{eqnarray}

As before, $u_i$, $u_j$, $u'_i$ and $u'_j$ are the amount
of money before and after the interaction of agent $i$ and $j$ respectively.

\item
The application of all of the above
``rules'' will be denoted as ``step''. 
These steps are successively and indefinitely repeated.

\end{enumerate}

The steady state PDF in the numerical 
simulations of this mapping are approximated by a Gamma distribution (Eq. \ref{eq:gamma2}),  
where $\langle u \rangle$ is the mean wealth of the ensemble of agents, $a$ is given by
Eq. \ref{eq:Gamma_a} and $n$ verifies the relationship

\begin{equation}
\label{eq:gamma:n}
 n=\frac{3 - 2 \omega}{2 \omega}.
\end{equation}

This function, together with a molecular dynamics simulation 
of the model are shown in Fig. \ref{fig:gamma1}.
Although this Gamma function apparently fits very well the numerical
simulation,
we will show here that
in fact it is just an approximation to the PDF
and it is only exact for the cases where $\omega = 1$ or $\omega = 1/2$.

Let us now obtain the analytical expression for the PDF. We will proceed
as before. Rule \ref{gamma1:rl:1} will deplete
the PDF for those particular values of $u$

\begin{equation}
\label{eq:gamma1:deltafu1}
\frac{\partial f(u_i)}{\partial t} \sim - f(u_i),
\end{equation}

and

\begin{equation}
\label{eq:gamma1:deltafu2}
\frac{\partial f(u_j)}{\partial t} \sim - f(u_j).
\end{equation}

From Eqs. \ref{eq:gamma1} we can derive the limits
of the variables. From the first of these equations (\ref{eq:gamma1})
we obtain the inequalities

\begin{equation}
 u_i < u'_i < u_i + \omega \; u_j,
\end{equation}

from which we see that the variable $u'_i$ is spanning an interval
of length $\omega \; u_j$. This interval is where the probability 
from $\epsilon$ spreads over,
so we need to divide by this factor. 
We can now split this inequation in the two following ones

\begin{eqnarray}
 u_i < u'_i \nonumber, \\
 u_j > (u'_i - u_i)/\omega,
\end{eqnarray}

and from the second equation of the mapping (\ref{eq:gamma1}) we obtain

\begin{eqnarray}
 u_j > u'_j \nonumber, \\
 u_j < u'_j/(1-\omega).
\end{eqnarray}

The increment of the PDF coming from the first equation of the mapping \ref{eq:gamma1}
will then be

\begin{equation}
\label{eq:gamma1:deltafup1}
 \frac{\partial f(u'_i)}{\partial t} \sim \int_{0}^{u'_i} \mathrm{d} u_i \int_{(u'_i - u_i)/\omega}^\infty 
\mathrm{d} u_j 
\frac{f(u_i) f(u_j)}{\omega u_j},
\end{equation}

and similarly it is obtained

\begin{equation}
\label{eq:gamma1:deltafup2}
 \frac{\partial f(u'_j)}{\partial t} \sim \int_{u'_j}^{u'_j/(1-\omega)} \mathrm{d} u_j \int_{0}^{\infty} 
\mathrm{d} u_i 
\frac{f(u_i) f(u_j)}{\omega u_j}.
\end{equation}

where the integration limits of $u_i$ go from $0$ to $\infty$ because there are no constraints
on this variable in the second of Eqs. \ref{eq:gamma1}. Since $f(u_i)$ is normalized
we can remove it from this expression,

\begin{equation}
\label{eq:gamma1:deltafup2_2}
 \frac{\partial f(u'_j)}{\partial t} \sim \int_{u'_j}^{u'_j/(1-\omega)} \mathrm{d} u_j  
\frac{f(u_j)}{\omega u_j}.
\end{equation}

The final expression (Eq. \ref{eq:formula1}) for the evolution of the PDF is

\begin{eqnarray}
\label{eq:gamma1:deltafu}
N \frac{\partial f(u)}{\partial t} &=& \nonumber \\
&-&  2 f(u) \nonumber \\ 
&+&  \int_{0}^{u} \mathrm{d} u_1 \int_{(u - u_1)/\omega}^\infty 
\mathrm{d} u_2 
\frac{f(u_1) f(u_2)}{\omega u_2} \nonumber \\
&+& \int_{u}^{u/(1-\omega)} \mathrm{d} u_2
\frac{f(u_2)}{\omega u_2},
\end{eqnarray}

where we have again written the dummy variable generically as $u$, $u_1$
and $u_2$ and $N$ is the normalization factor of $\partial f(u)/\partial t$.

The steady state PDF can now be readily derived from it 
by setting $\partial f(u)/\partial t=0$,

\begin{eqnarray}
\label{eq:gamma1:final}
f_{eq}(u) = \nonumber \\ 
&+& \frac{1}{2} \int_{0}^{u} \mathrm{d} u_1 \int_{(u - u_1)/\omega}^\infty 
\mathrm{d} u_2 
\frac{f_{eq}(u_1) f_{eq}(u_2)}{\omega u_2} \nonumber \\
&+& \frac{1}{2} \int_{u}^{u/(1-\omega)} \mathrm{d} u_2
\frac{f_{eq}(u_2)}{\omega u_2}.
\end{eqnarray}

We can solve this integro-differential equation 
iteratively by substituting $f(u)$
with the known approximate solution \ref{eq:gamma2} 
in the right hand side of Eq. \ref{eq:gamma1:final} to give
a second order approximation in the left hand side. 
Usually one iteration is enough to obtain
a more precise solution. 
The solution of this procedure is shown in Fig. \ref{fig:gamma1} 
as a red dashed line.
We can see in this figure that this solution is indistinguishable 
from the first approximation in Eq. \ref{eq:gamma2}
shown in Fig. \ref{fig:gamma1} as a solid blue line,
but they are not exactly the same, as we will demonstrate now.

\begin{figure}
\includegraphics[angle=0,width=0.9\columnwidth]{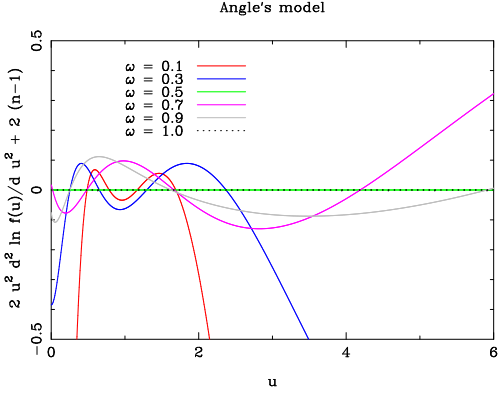}
\caption{\label{fig:gamma1:derivative} Plot of
$2 u^2 \; \frac{\mathrm{d}^2 \ln f(u)}{\mathrm{d} u^2}  + 2 (n-1)$,
where the first term has been
calculated using Eq. \ref{eq:gamma1:derivative}, 
as a function of $u$. It is observed that the Gamma
function is an exact solution only for $\omega$ equal $1$ or $1/2$.
For these calculations, we take $<u>=1$.}
\end{figure}

If the Gamma function (Eq. \ref{eq:gamma2}) is indeed 
an exact solution of Eq. \ref{eq:gamma1:final},
the second derivative of the logarithm of both expressions should also match.
The second derivative of the logarithm of the 
Gamma function (Eq. \ref{eq:gamma2}) is

\begin{equation}
 \frac{\mathrm{d}^2 \ln f_{eq}(u)}{\mathrm{d} u^2} = - \frac{(n-1)}{u^2}.
\end{equation}

If we now multiply by $2 u^2$ we obtain a constant,

\begin{equation}
\label{eq:gamma1:derivative_gamma}
 2 \; u^2 \; \frac{\mathrm{d}^2 \ln f_{eq}(u)}{\mathrm{d} u^2} = - 2 (n-1).
\end{equation}

Repeating this same calculation in Eq. \ref{eq:gamma1:final} 
by  introducing the 
Gamma function (Eq. \ref{eq:gamma2}) in its right hand side,
and doing some more elaborate calculations, we find the result,

\begin{eqnarray}
&& 2 u^2 \; \frac{\mathrm{d}^2 \ln f_{eq}(u)}{\mathrm{d} u^2} = \nonumber \\
&& -\int_{0}^{u} \mathrm{d} u_1 (n-1) \frac{a (u - u_1)^{(n-2)}  u_1^{(n-1)} }{ \omega^n u^{(n-2)} }
\mathrm{e}^{ \frac{n}{<u>} \frac{\omega -1}{\omega} u_1 } \nonumber \\
&& -  \frac{\mathrm{e}^{-\frac{n}{<u>} \frac{\omega \; u}{1-\omega}}}{\omega (1-\omega)^{(n-1)}}
      \left( 1 + \frac{n}{<u>} \frac{\omega \; u}{1-\omega} \right) \nonumber \\
 &&     +\frac{1}{\omega}.
\end{eqnarray}

The integral can be solved analytically to finally give,

\begin{eqnarray}
\label{eq:gamma1:derivative}
&& 2 u^2 \; \frac{\mathrm{d}^2 \ln f_{eq}(u)}{\mathrm{d} u^2} = \nonumber \\
&&-\frac{a (n-1) u^{n} \Gamma(n-1) \Gamma(n) }{ \omega^{n} \Gamma(2n-1)} 
  ~_{1}F_{1} \left(n,2n-1, \frac{n}{<u>} \frac{\omega -1}{\omega} u \right) \nonumber \\
 &&- \frac{\mathrm{e}^{-\frac{n}{<u>} \frac{\omega \; u}{1-\omega}}}{\omega (1-\omega)^{(n-1)}}
       \left( 1 + \frac{n}{<u>} \frac{\omega \; u}{1-\omega} \right) \nonumber \\
  &&   +  \frac{1}{\omega},
\end{eqnarray}

where $~_1F_1$ is the confluent hypergeometric function.
If $f(u)$ is really a Gamma function, the subtraction of
Eq. \ref{eq:gamma1:derivative} from the expected result 
in Eq. \ref{eq:gamma1:derivative_gamma} should give zero.
So the question to be answered in this case is:

\begin{equation}
\label{eq:gamma1:zero}
 2 u^2 \; \frac{\mathrm{d}^2 \ln f_{eq}(u)}{\mathrm{d} u^2}  + 2 (n-1) = 0 \; ?
\end{equation}

If this result is not zero, it indicates that $f_{eq}(u)$ 
is not exactly a Gamma function.
Results of these calculations are shown in Fig. \ref{fig:gamma1:derivative} 
for different values of $\omega$.
As it can be seen, the only results that makes this difference equal to zero 
over the whole domain of $u$ are 
$\omega = 1$ or $1/2$. Let us prove it.

If $\omega=1$ then $n=1/2$ (Eq. \ref{eq:gamma:n}) 
and the first two terms of the right hand side
of Eq. \ref{eq:gamma1:derivative} are zero,
leaving Eq. \ref{eq:gamma1:zero} as

\begin{equation}
 \frac{1}{\omega} + 2 (n-1) = 0,
\end{equation}

which we can readily verify is identically zero.

If $\omega=1/2$ then $n=2$ and the hypergeometric function can be simplified to

\begin{equation}
 \frac{~_1F_1}{\Gamma(2 n-1)} = \frac{\mathrm{e}^{-2\frac{u}{<u>}} (-1 - 2\frac{u}{<u>}) }
 {4 \frac{u^2}{<u>^2}} + \frac{1}{4 u^2/<u>^2}.
\end{equation}

The first term of the right hand side of Eq. \ref{eq:gamma1:derivative} 
is then

\begin{equation}
 -4 \mathrm{e}^{-2\frac{u}{<u>}} \left( -1 - 2 \frac{u}{<u>}\right) - 4.
\end{equation}

Inserting the values of $\omega$ and $n$, we can readily verify that 
the second term of the right hand side of Eq. \ref{eq:gamma1:derivative} is
\begin{equation}
 -4 \mathrm{e}^{-2\frac{u}{<u>}} \left( 1 + 2 \frac{u}{<u>}\right).
\end{equation}

Adding both together with the last term of 
Eq. \ref{eq:gamma1:derivative} and introducing
this result in Eq. \ref{eq:gamma1:zero} gives

\begin{equation}
 -4 + \frac{1}{\omega} + 2 (n-1) = 0,
\end{equation}

which we can verify is true.

To calculate the relaxation times we proceed as before.
We can simplify the evolution equation
(Eq. \ref{eq:gamma1:deltafu}) close to equilibrium to

\begin{equation}
 N \frac{\partial f(u)}{\partial t} \simeq -  2 f(u) +  2 f_{eq}(u).
\end{equation}

And again, the relaxation time will be $\tau = N/2$ collisions.

\section{Conclusion}

A step by step guide to derive the dynamical evolution equations
of the PDFs of economic models involving money exchange
between interacting agents, or for that matter,
any other type of similar mappings, has been explained.
The equilibrium distribution can be found by exploring its
stationary solutions.
This leads to an integro-differential
equation which can be solved analytically in some cases and
numerically in others.

Dragulescu and Yakovenko's mapping \cite{dragulescu00}
can be solved analytically giving exactly an exponential
distribution (see Fig. \ref{fig:exponential}).
Chakraborti and Chakrabarti's model \cite{chakraborti00}
can be approximated by a Gamma function, but this result
is not precise, especially for low values of the savings
parameter, $\lambda$. In this case the numerical solution
of the integro-differential equation provides a better fit
to the molecular dynamics numerical simulations (Fig. \ref{fig:chakra}).
Angle-like's model \cite{angle83} steady state PDF can be approximated
very well by a Gamma function (Fig. \ref{fig:gamma1}), 
but according to the more exact
integro-differential equations derived here, 
it is not an exact solution to the problem except
for values of the money exchange parameter, $\omega$, of
$1$ or $1/2$ (Fig. \ref{fig:gamma1:derivative}).

This technique should prove useful to get insights
into the type of PDFs we can expect from a particular
mapping, coming this mapping from an economic exchange model
or from a gas with colliding particles. The exchange rules,
in the case of economic models,
or the collision specifications, in the case of
gases, will determine which kind of PDFs the system
will reach in the steady state.
It is interesting to draw this parallelism, because in the economic
case, and with the models shown here, we have mappings which
are not symmetric in time (the equations to deplete and increment
the PDFs in one time step are totally different) which give rise
to distributions of the Gamma function family.
In the case of gases, and in particular with the
Boltzmann integro-differential equation, the collisions
are symmetric in time (the equations to deplete and increment
the PDFs in one time step are symmetrical), giving rise to 
many interesting properties, like a Gaussian function 
for the steady state PDF
and the possibility to prove analytically Boltzmann's H theorem.

It should be noted that we are obtaining the
time evolution equations of the PDFs, and not only the stationary
solutions. This should allow the study of the dynamical
evolution of these systems, something which is key to economics,
which as we know is not static. This has been illustrated in this
paper by calculating the relaxation times of these system,
but many other applications could be devised.

\begin{acknowledgments}
The figures in this paper have been prepared using the numerical
programming language PDL (http://pdl.perl.org).
\end{acknowledgments}

\bibliography{anagas4}{}

\end{document}